\documentclass{mem}
\usepackage{txfonts}\usepackage{balance}
\usepackage{graphicx}
\idline{75}{1}
\begin{document}
\def\teff{$T\rm_{eff }$}
\def\kms{$\mathrm {km s}^{-1}$}

\title{
Cluster Turbulence: Simulation Insights
}


   \author{T. W. Jones
          \inst{1,2},
             David H. Porter\inst{2},
          Dongsu Ryu\inst{3}
         \and
	 Jungyeon Cho\inst{3}
	  }
   \institute{Department of Astronomy, University of Minnesota,
              Minneapolis, MN 55455, USA\\
          \and
              University of Minnesota Supercomputing Institute,
	      Minneapolis, MN 55455, USA\\
	      \email{twj@astro.umn.edu, porter@msi.umn.edu }
          \and
	      Department of Astronomy and Space Science, 
	      Chungnam National University, Daejun 305-764, Korea\\
	      \email{ryu@canopus.chungnam.ac.kr, cho@canopus.cnu.ac.kr}
}

  \offprints{T. Jones}

\authorrunning{Jones }

\titlerunning{Cluster Tubulence}

\abstract{
   Cluster media are dynamical, not static;
   observational evidence suggests they are turbulent.
   High-resolution simulations of the intracluster media (ICMs)
   and of idealized, similar media help us understand
   the complex physics and astrophysics involved.
   We present a brief overview of the physics behind ICM turbulence
   and outline the processes that control its development.
   High-resolution, compressible, 
   isothermal MHD simulations are used to illustrate important
   dynamical properties of turbulence that develops in media with
   initially very weak magnetic fields. The simulations follow
   the growth of magnetic fields and reproduce the characteristics of
   turbulence. These results are also compared with
   full cluster simulations that have examined the
   properties of ICM turbulence.
\keywords{galaxy clusters -- magnetohydrodynamics -- turbulence}
}
\maketitle{}

\section{Introduction}

   Observation and theory have revealed 
   intracluster media (ICMs) to be very dynamic environments with
   active ``weather'' driven by a host of activities such as mergers, accretion,
   AGNs, galactic winds and instabilities.
   These drivers are common and cause the ICMs
   to be criss-crossed by large-scaled, complex flows that generate shocks,
   contact discontinuities (aka ``cold fronts'') and bulk shear. Inevitably
   such flows should lead to turbulence in the ICMs, an outcome
   supported by growing observational evidence.
   These include, for example, substantial ICM random velocities in Perseus
   reducing resonance scattering in the 6.7 keV iron line \cite{chur04},
   evidence for thermal ICM pressure fluctuations in the Coma cluster
   \cite{schu04}, patchy Faraday rotation measure distributions in several
   clusters \cite{bona10} and the absence of large scale polarization in
   cluster radio halos \cite{kim90}, suggesting disordered magnetic fields.

   Turbulence in clusters is important to understand for many reasons.
   Turbulent pressure helps support the ICM, so relevant
   cluster mass measures. Turbulence transports 
   entropy, metals and cosmic rays, all
   important cluster diagnostics. It transports
   and amplifies magnetic fields, which in turn control ICM
   viscosity, resistivity and thermal conductivity, as well as the propagation 
   and acceleration of cosmic rays.
   The literature on turbulence is extensive including
   excellent reviews on MHD turbulence, which is
   most relevant to the ICM (e.g., \cite{brand05}). Here we make a few observations
   pertinent to this meeting.


\section{Origins of Cluster Turbulence}

Turbulence describes motions possessing significant 
random velocities. This random velocity can
include both compressible ($\nabla\cdot \vec{u} \ne 0$) and vortical, or 
solenoidal ($\vec{\omega} = \nabla\times \vec{u} \ne 0$) components.
In ICM circumstances, which usually involve mostly subsonic
flows, the vortical component ordinarily dominates. Thus, 
understanding this turbulence begins with an 
identification of the sources of vorticity and the manner in which vorticity evolves. 

Euler's equation 
for an ideal fluid can be expressed in terms of vorticity \cite{landau},
\begin{equation}
\frac{\partial \vec{\omega}}{\partial t} = \nabla\times(\vec{u}\times\vec{\omega}) + \frac{1}{\rho^2}\nabla\rho\times \nabla P. 
\label{eq:vort}
\end{equation}
This can be rewritten as a conservation law for the integrated vorticity
through a surface element moving with the fluid, or
by way of Stokes' Theorem a conservation of circulation around 
that surface element,
\begin{equation}
\frac{d}{dt}\int\vec{\omega}\cdot d\vec{a} = \frac{d}{dt}\oint \vec{u}\cdot d\vec{l} = \int \frac{1}{\rho^2}(\nabla\rho\times\nabla P)\cdot d\vec{a},
\label{eq:vcons}
\end{equation}
where $d /dt$ is the Lagrangian time derivative.

Equation \ref{eq:vcons} shows that in ideal, barotropic flows, where the pressure
depends only on density ($\nabla\rho\times \nabla P = 0$), a fluid element's net vorticity 
is conserved.
Local values of $\vec{\omega}$ still change, of course, without
vorticity source terms. 
For instance, when
an incompressible flow element's cross section is decreased as it is stretched,
$\vec{\omega}$ will increase.
This vortex stretching (how tornados form) is central to 
evolution of turbulent flows. 

These vorticity properties are analogous to
magnetic flux conservation familiar to all astrophysicists. In fact, the
magnetic induction equation in ideal MHD is the same as the barotropic
form of equation \ref{eq:vort} with the substitution $\vec{\omega} \rightarrow \vec{B}$.
The vortex stretching analogy applied to magnetic fields
means that stretched magnetic flux tubes enhance local 
fields. The local B of an incompressible flux tube varies in proportion to the length of the tube.
Flux tube stretching is, in fact, at the core of
the turbulence dynamo, or small scale dynamo that amplifies weak magnetic fields inside
turbulent conducting media. Vortical
turbulent motions statistically extend the length of a fluid element over time,
causing both vorticity and magnetic field to be locally enhanced while
conserving total circulation and magnetic flux in an ideal fluid.
The root-mean-square (RMS) values of both vorticity and magnetic field intensity grow in this
way, even in the absence of source terms.

On the other hand, vorticity can be generated at the curved surface
of shocks in and around clusters (e.g., \cite{ryu03,kang07,vazza09}) and by the
baroclinity of flows.
For uniform upstream flow, the vorticity behind a curved
shock surface is
\begin{equation}
{\vec \omega}_{\rm cs} = \frac{(\rho_2 - \rho_1)^2}{\rho_1 \rho_2}
K {\vec U}_1 \times {\hat n},
\label{eq:shock_vort_gen}
\end{equation}
with $\rho_1$ and $\rho_2$ the upstream and downstream gas
densities, ${\vec U}_1$ the upstream flow velocity
in the shock rest frame, $K$ the shock surface curvature tensor,
and ${\hat n}$ the shock normal unit vector.
If isopycnic (constant density)
and isobaric surfaces are not coincident,
vorticity is generated
according to equation \ref{eq:vcons}.
If we let $P = S\rho^{\gamma}$, where $S$ is a proxy for
entropy in a $\gamma$-law gas, we see that the baroclinity is introduced by 
shock induced entropy variations.
See \cite{ryu08} for more discussion of this issue.
AGNs and galactic winds also add shear to ICMs, so equivalently,
vorticity ($\omega \sim \Delta u/\delta$, where $\delta$ is the
thickness of the shear layer).

Subsequently, the energy in the vortical flows cascades down to
smaller scales and turbulence develops,
provided the viscous dissipation scale, $l_{visc}$, is smaller
than the ``driving scale'', $L_d$, for the flow.
The driving scale, $L_d$,
is generally comparable to such things as the curvature radius of a shock, the
size of a substructure core, or the scale of an AGN or galactic
outflow. These likely range from $\sim10\rm{s}$ of kpc to 
$\sim 100\rm{s}$ of kpc.

The appropriate viscous dissipation scales in ICMs remain uncertain.
The media are hot and very diffuse plasmas, so Coulomb collisions
are ineffective.
The associated mean free path, $\lambda_{Coul} \sim
1~\rm{kpc}~T_{keV}^{5/2}/(n_{-3}u_{th,100})$, 
ranges from $\sim 10$s of pc 
to $\sim 10$s of kpc, depending on the cluster circumstances.
Here $T_{keV}$ is the ICM temperature
in keV, $n_{-3}$ is the density in $10^{-3}\rm{cm}^{-3}$ and $u_{th,100}$ is
the thermal velocity of ions in $100~\rm{km/sec}$.
The corresponding kinetic viscosity, $\nu \sim  u_{th}\lambda_{Coul}$, is very large,
and the Reynolds number, $Re \sim L_d U / \nu\sim \rm{few~}\times 10$ , 
with $U \sim u_{th}\sim \rm{few~to~} 10s$, the velocity at
the driving scale.
Hence, the viscous dissipation scale due to Coulomb scattering alone,
$l_{visc} \sim L_d /Re^{3/4}$, would range from fractions of a kpc
in cooling cores to several 10s of kpc in  hot ICMs.
On the other hand, it has been suggested that plasmas threaded with weak magnetic
fields, like the ones in the ICMs, are subject to gyro-scale instabilities, such
firehose and mirror instabilities; then,
the scattering of particles with the resulting fluctuations could reduce the
particle mean-free path, so also the viscous dissipation scale (\cite{sche06}).
The detailed picture is still very uncertain.
Nevertheless, in our discussion below we assume
the physical dissipation scale is at least as small as the
effective dissipation scales of the simulations, so
of order the grid resolution. In our simulations, the
grid resolution would be $\sim 1$ kpc or so (see \S 3),
while for full cluster simulations it is at least several kpc.

The resistive dissipation scales, $l_{res}$, in the ICMs
are also uncertain.
In a turbulent flow with $\eta \ll \nu$, it is $l_{res} \sim L_d /Rm^{1/2}$, where $Rm \sim L_d U / \eta$ is
the magnetic Reynolds number, and $\eta$ is the resistivity.
It is likely that the magnetic Prandtl number $P_{r,M} \equiv Rm/Re = \nu/\eta \ga 1$ in the ICMs.
For instance, viscosity and resistivity due exclusively to Coulomb
scattering would lead to $P_{r,M} \gg 1$ (e.g., \cite{sp78}).
In the simulations reported here, both the viscosity and resistivity have
numerical origins with
dissipation scales of order the grid resolution; thus, $P_{r,M} \sim 1$. 
Most simulations, including most full cluster simulations, effectively have $P_{r,M} \sim 1$ too.
Strictly speaking, the simulations with $P_{r,M} \sim 1$ should be valid only in,
and so applied only to the scales of $l \ga l_{visc}$ and $l \ga l_{res}$.
For the scales of $l_{res} \la l \la l_{visc}$, simulations with
large Prandtl number are required.
It is, however, hard to reproduce large Prandtl number turbulence with sufficient
viscous and resistive inertial ranges (e.g., \cite{schek04}).

\section{Simulation of ICM-like Turbulence}

It is not simple to isolate turbulence from generally complex motions in real or
simulated clusters. As a complementary exercise
we have initiated a high-resolution simulation study of the
evolution and saturation of driven MHD turbulence in periodic computational domains 
that resemble ICMs.
Since cluster media, while clearly
magnetized, are not energetically dominated by
magnetic fields, we focus on 
turbulence developed with initially very weak magnetic fields. The full study considers
both compressible and incompressible fluids and ideal
and nonideal media with a range of magnetic Prandtl numbers. We report here some
initial results of 
simulations of ideal, compressible MHD turbulence in isothermal media.
The simulations used an isothermal ideal MHD code,
which is a version updated from that of \cite{kim01} for massive parallelization.
Results here are from simulations with $1024^3$ and $2048^3$
grid zones (with the typical cluster sizes $\sim 1$ Mpc, the grid resolution
would be $\sim 1$ kpc).
The two simulations are very consistent; the higher resolution calculation
produces slightly stronger magnetic fields at saturation.

Initially the medium has uniform density, $\rho = 1$, with gas pressure, $P = 1$,
so that the sound speed, $c_s = 1$.
The initially magnetic field is uniform and very weak, with $\beta = P/P_B = 10^6$. 
The cubic box has dimensions, $L_0 = L_x = L_y = L_z = 10$ with periodic
boundaries.
Turbulence is driven by solenoidal velocity forcing,
drawn from a Gaussian random field
determined with a power spectrum $P_k \propto k^6 \exp(-8 k / k_{peak})$,
where $k_{peak} = 2 k_0$ $(k_0 = 2 \pi / L_0)$, and added at every
$\Delta t = 0.01 L/c_s$.
The power spectrum peaks around $k_d \approx 1.5 k_0$, or
around a scale, $L_d \approx 2/3 L_0$.
The amplitude of the perturbations is tuned so that
$u_{RMS} \sim 0.5$ or
$M_s \equiv u_{RMS} /c_s \sim 0.5$ at saturation, close to
what resuilts in full cluster simulations (e.g., \cite{nagai07,ryu08,vazza09},
and see also \S 4).

\begin{figure}
   \centering
   \includegraphics[width = 0.50\textwidth]{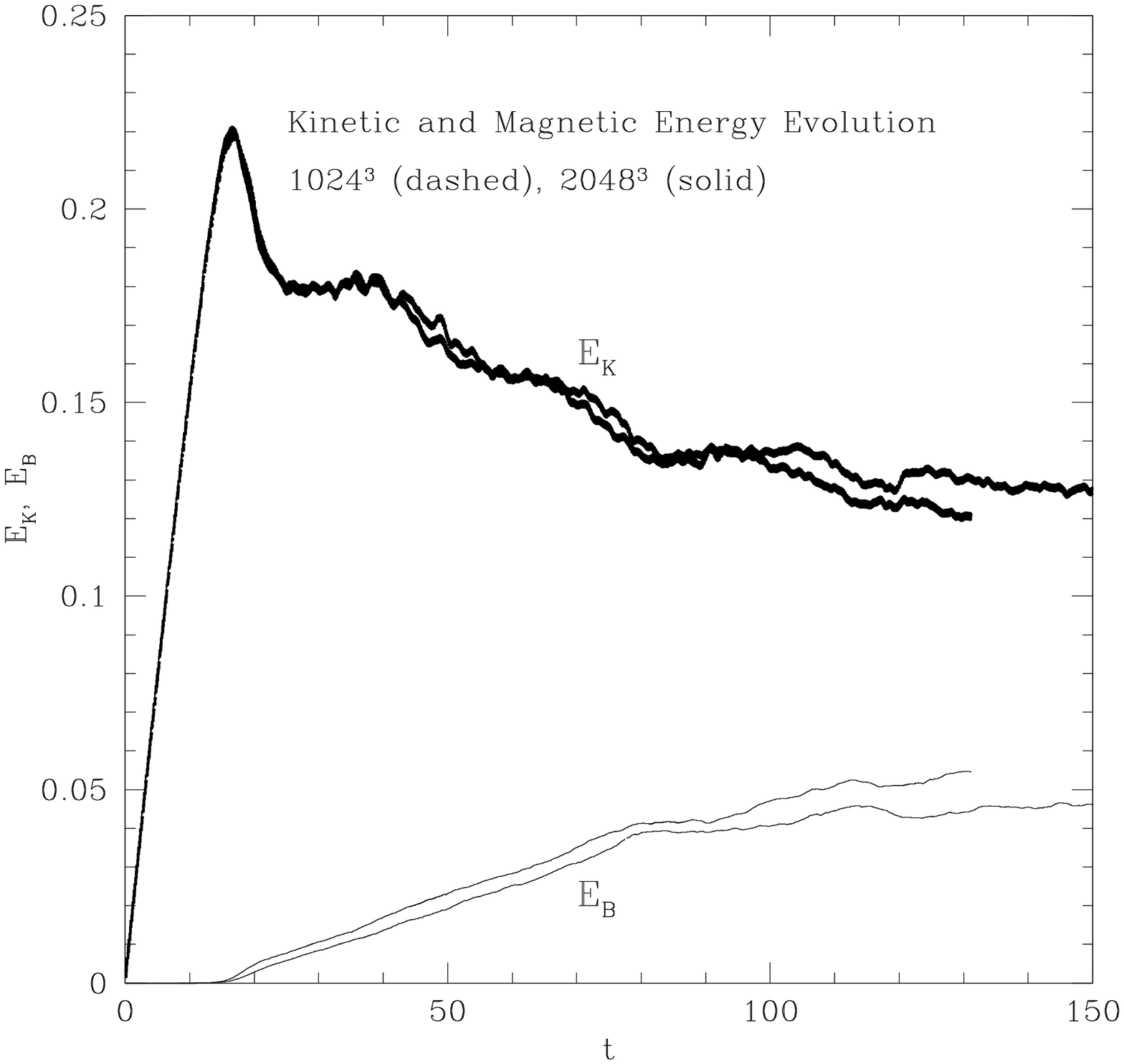}
   \includegraphics[width = 0.50\textwidth]{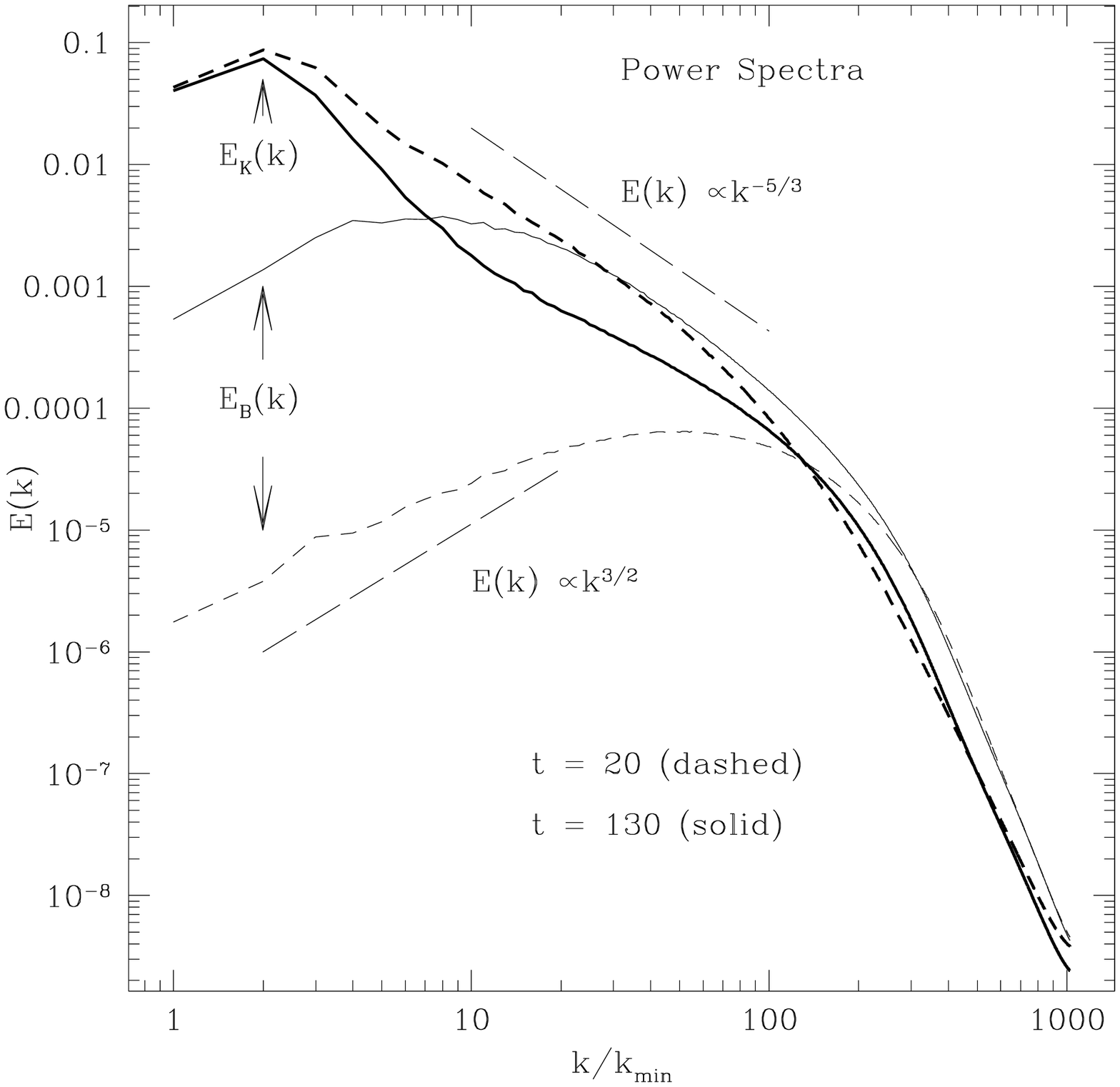}
   \caption{Top: Evolution of kinetic, $E_K$, and magnetic, $E_B$,
   energies in simulations of ideal, compressible MHD turbulence
   for two grid resolutions. Bottom: Power spectra, $E(k)$,
   of kinetic and magnetic energies at $t = 20$
   and $t=130$ in the $2048^3$ zone turbulence.}
              \label{lineplots}%
    \end{figure}

Our setup gives a characteristic time scale, 
$t_d = L_d/ u_{RMS} \approx 15$.  In that time the largest eddies should
spawn something resembling turbulent behavior.
Indeed Fig. \ref{lineplots}
shows that the mean turbulent kinetic energy density, $E_K$, grows 
and peaks at $t \sim t_d$, with
a value corresponding to $u_{RMS} \sim 2/3$.
Subsequently, $E_K$ slowly declines as
the mean magnetic energy density, $E_B$ grows. The kinetic energy power spectrum, $E_K(k)$,
at $t = 20$
also shown in Fig. \ref{lineplots}, exhibits a peak at $k/k_0 \sim 2$, near the driving scale,
and takes a Kolmogorov-like, inertial form, 
$E(k) \propto k^{-5/3}$, for $k/k_0 \la 50$. 
By this time energy has cascaded from the driving scales far enough that the motions, with
negligible magnetic backreaction, are reasonably described as classical, hydrodynamical
turbulence over a modest range of scales.

   \begin{figure}
   \centering
   \includegraphics[width = 0.50\textwidth]{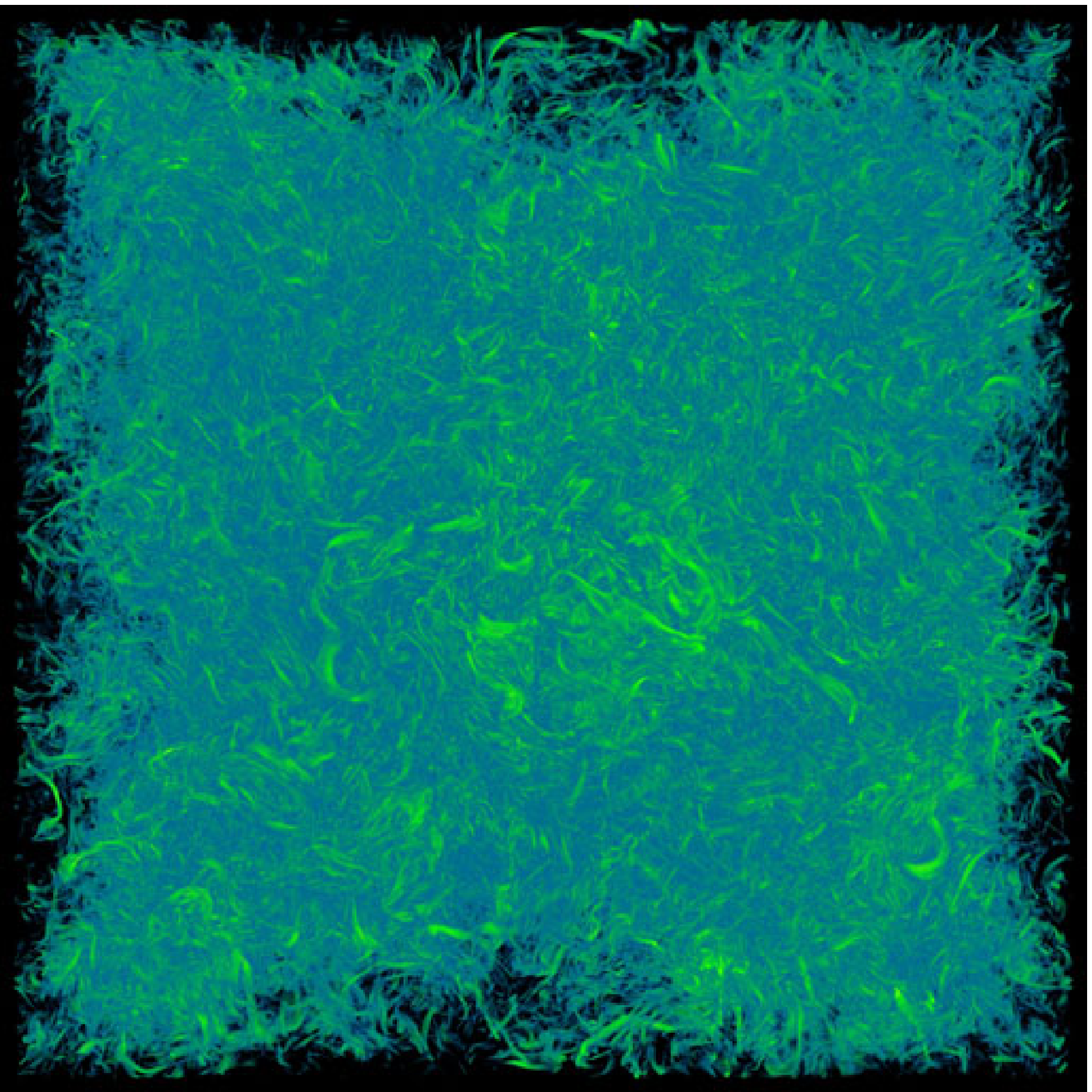}
   \includegraphics[width = 0.50\textwidth]{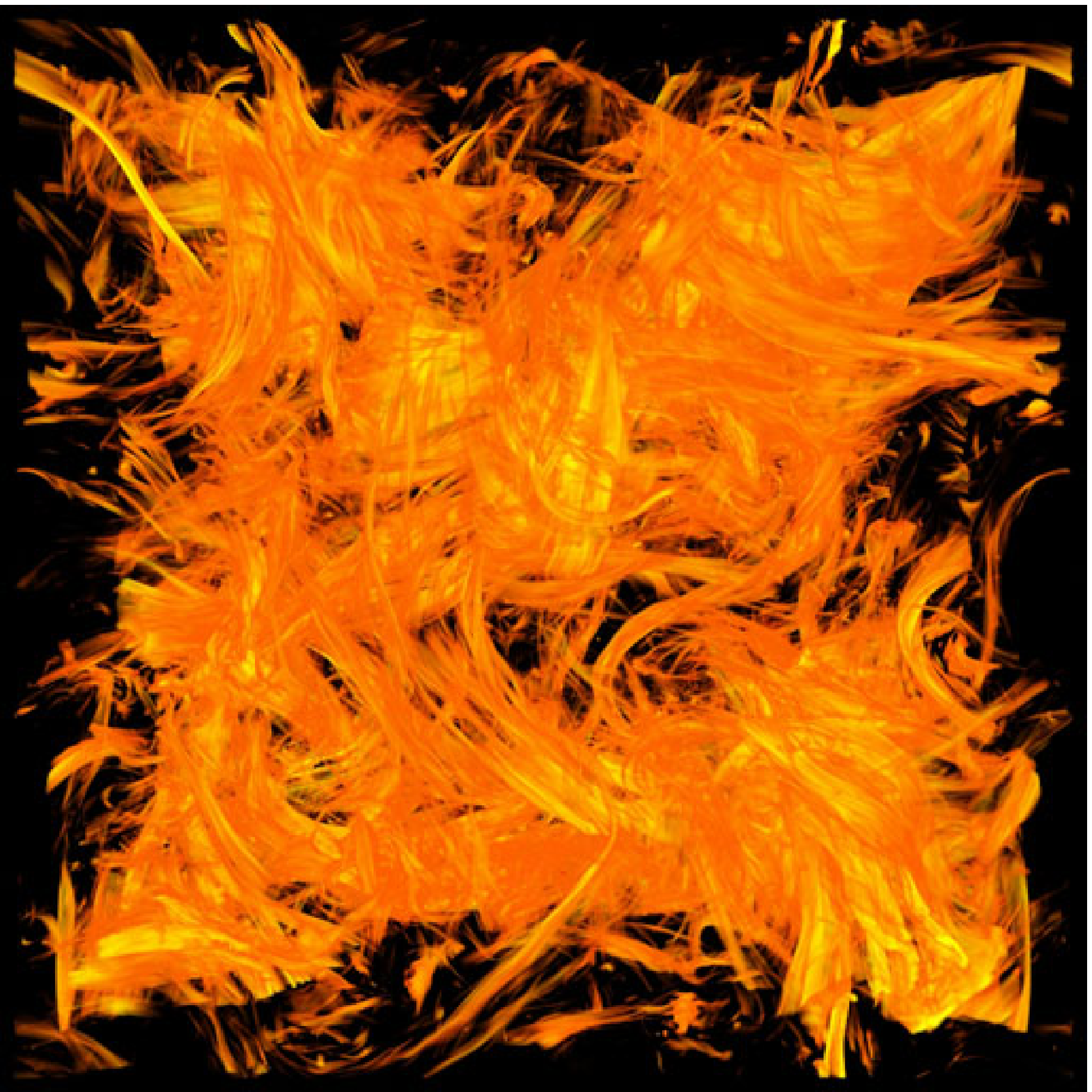}
   \caption{Magnetic energy density distributions
   in MHD turbulence. Top: Log($E_B$)
   at $t = 20$ in $2048^3$ simulation.
   Bottom: Log($E_B$)
   at $t = 130$. Higher tones at the late time reflect stronger fields.}
              \label{images}%
    \end{figure}

Turbulent fluid motions stretch and intensify vorticity and magnetic flux, leading
to development of a chaotic sea of vortex and magnetic filaments.
This is illustrated for
the magnetic field at $t = 20$ in the top of Fig. \ref{images}. 
The magnetic filaments in this image have characteristic lengths of a few \% of $L_0$.
This agrees with the fact that the magnetic power spectrum, $E_B(k)$,
peaks for $k_{peak}/k_0 \sim 50$ at this time. The transverse dimensions of
the filaments, an order of magnitude smaller at this time, seem to correspond to
the dissipation scale.

As the magnetic field intensifies, both vorticity and magnetic energies undergo inverse
cascades from small to large scales, with the coherence lengths of their filaments growing
accordingly.
This is evident for the magnetic field in the power spectrum changes in Fig. \ref{lineplots}.
The inverse cascade of magnetic energy can be understood as follows.
The field is wrapped more quickly around smaller scale eddies,
because the eddy turn over time varies as $t_l \propto l^{2/3}$. 
Maxwell stresses, $\propto (\nabla\times B)\times B$, then, feed back on the kinetic turbulence,
causing significant modifications in the fluid motions, thus saturating
the magnetic field growth on a given scale, $l$, when $E_B(l) \sim E_K(l)$. 
Since the turbulent kinetic energy on a scale $E_K(l) \propto l^{2/3}$,
the saturation scale of the magnetic turbulence should evolve over time
as $l_B \propto t^{3/2}$, while the magnetic energy grows
linearly with time, both consistent with Fig. \ref{lineplots}.
Eventually, as $l_B$ approaches the outer scale of the kinetic turbulence, $L_d$,
the scalings break down and turbulence reaches saturation where the ratio of the total
magnetic to kinetic energy is $E_B/E_K \sim 2/3$ (see also, e.g.,
\cite{schek04, ryu08, cho09a,cho09b}).

Finally, we emphasize an interesting topological transformation in
the flow structure as turbulence proceeds through the linear growth to the saturation
stage.
Fig. \ref{images} displays the 
different topologies of the magnetic flux structures at
$t = 20$ and $t = 130$. At the earlier time the field is organized into individual
filaments. At the later time those filaments have evolved into striated, ribbon-like forms
(see also \cite{schek04}).
Close examination reveals the ribbons to be laminated, with
magnetic field and vorticity interleaved through each cross section
on scales of the order the dissipation length.
In hydrodynamical turbulence such ribbons would be unstable, but the interplay of
vorticity and magnetic field seems to stabilize them in MHD turbulence.

The $t = 130$ image in Fig. \ref{images} also highlights the important
fact that the
magnetic field in MHD tubulence is highly intermittent. Relatively strong field
ribbons wrap around large shear leaving weak field cavities. 
Such structures are quite distinct from what one obtains,
for example, if they construct a magnetic field distribution out of
a Gaussian random variate, even if the outcome is a magnetic field
distribution having exactly the same power spectrum, as 
illustrated very nicely in previous work by \cite{waelk09}.

\section{Comparison to Cluster Simulations}

Recent high-resolution cluster formation
simulations have been analyzed for information about 
properties of ICM turbulence and its evolution. A few of these
simulations include MHD e.g., \cite{donn09, ruszk10, xu10}.
With or without magnetic fields, an initial challenge is identifying true
turbulence in generally complex flows, especially during and after
merging activity. Purely solenoidal motions in simulations in a periodic
domain can be
cleanly isolated using Fourier transforms (e.g., \cite{ryu08}). Otherwise
some kind of spatial filtering is needed that analyses the
motions only up to a maximum scale, such as the core radius of the
cluster (e.g., \cite{dolag05}). We mention here a few findings of
general significance in this context.

Several studies have emphasized turbulence generation from
shocks in and around clusters, especially coming from structure formation generally, and
merger activity specifically
(e.g., \cite{kang07,ryu08,vazza09}). Several studies found that
turbulent energy in post-merger ICMs can be comparable to, although somewhat
smaller than, the thermal energy;
it commonly reaches levels $E_K \sim 1/10 - 1/4~E_T$ (\cite{nagai07,ryu08,vazza09}), 
similar to our simulations discussed in \S 3. Thus, turbulent pressure
may impact on hydrostatic equilibrium-based mass estimates.
Simulations also suggest that the relative
turbulent pressure support is greater towards cluster outskirts
(e.g., \cite{ryu08,lau09}).
The turbulence formed in full cluster
simulations does not have a sufficiently wide inertial range to evolve a
true Kolmogorov power spectrum. Within that limitation, however, the
reported kinetic energy power spectra are consistent with expectations. 
Several studies demonstrated that cluster turbulence
can amplify magnetic fields to at least $\mu$Gauss levels (e.g., \cite{donn09, xu10}).
This amounts to $\sim $a \% or so of the thermal
pressure and $\sim$10 \% or so of the kinetic turbulent
pressure. From simulations such as we reported here, the time to reach 
equipartition ($E_B \sim 2/3 E_K$) from an initially weak field is many
driving scale times. It is not surprising
that the fields produced in clusters are well below those levels. By the same token
the magnetic field power spectra may peak well 
below the driving scales. The magnetic power spectrum reported from
cluster formation simulations using magnetic fields seeded by AGNs in \cite{xu10}, for
example, is consistent with this expectation.

Turbulence decay generally takes
several eddy times on the driving scale  once the driving ends.
In clusters we expect decay times, $\tau_d \ga L_{d,100}/\Delta u_{100}$Gyr,
where $L_{d,100}$ in 100 kpc. This is consistent with turbulence decay times
$\sim$ Gyr reported in cluster formation simulations (e.g., \cite{paul10})

\section{Conclusions}

Processes such as shocks and outflows are likely to drive turbulence in
ICMs. The detailed physics is difficult to model analytically, but
simulations allow us to explore it
in some detail. Magnetic fields are integral
ingredients of both the microphysics of ICM transport
properties and essential players in the large scale dynamics. Simulations
are revealing important
insights into the character of the turbulence, including properties of
magnetic fields.

\begin{acknowledgements}
This work was supported in part
by the US National Science Foundation through grant AST 0908668,
National Research Foundation of Korea through grant 2007-0093860.
\end{acknowledgements}

\bibliographystyle{aa}

\begin{thebibliography}{}


  \bibitem[Bonafede etal 2010]{bona10} Bonafede, A., Feretti, L. Murgia, M., Govoni, F.,
  Giovannini, G., Dallacasa, D., Dolag, K. \& Taylor, G. B. 2010,
  A\&A, 513, 30
  \bibitem[Brandenburg \& Subramanian 2005]{brand05} Brandenburg, A. \& Subramanian, K.
  2005,  Phys. Rept., 417, 1
  MNRAS, 378, 245
  \bibitem[Cho etal 2009]{cho09a} Cho, J., Vishniac, E. T., Beresnyak, A., Lazarian, A. 
  \& Ryu, D. 2009, ApJ, 693, 1449
  \bibitem[Cho \& Ryu 2009]{cho09b} Cho, J. \& Ryu, D. 2009,
  ApJL, 705, L90
  \bibitem[Churazov etal 2004]{chur04} Churazov, E., Forman, W., Jones, C., Sunyaev, R.
  \& Bohringer, H. 2004, MNRAS, 347, 29
  \bibitem[Dolag etal 2005]{dolag05} Dolag, K., Vazza, F., Brunetti, G. \& Tormen, G.
  2005, MNRAS, 364, 753
  \bibitem[Donnert etal 2009]{donn09} Donnert, J., Dolag, K, Lesch, H. \& Muller, E. 2009,
  MNRAS, 392, 519
  \bibitem[Kang etal 2007]{kang07} Kang, H., Ryu, D., Cen, R. \& Ostriker, J. P. 2007,
  ApJ, 669, 729 
  \bibitem[Kim etal 2001]{kim01} Kim, J., Ryu, D., Jones, T. W. \& Hong, S. S. 2001,
  JKAS, 34, 281
  \bibitem[Kim etal 1990]{kim90} Kim, K.-T., Kronberg, P. P., Dewdney, P. E.
  \& Landecker, T. L. 1990, ApJ, 355, 29
  \bibitem[Landau \& Lifshitz 1987]{landau} Landau, L. D. \& Lifshitz ,E. M. 1987,
  Fluid Mechanics, 2nd Ed. (Pergamon, Oxford)
  \bibitem[Lau etal 2009]{lau09} Lau, E. T., Kravtsov, A. V. \& Nagai, D. 2009,
  ApJ, 705, 1129
  \bibitem[Nagai etal 2007]{nagai07} Nagai, D., Vikhlinin, A., \& Kravtsov, A. V.
  2007, ApJ, 655, 98
  \bibitem[Paul etal 2010]{paul10} Paul, S., Iapichino, L., Miniati, F., Bagchi, J.
  \& Mannheim, K. 2010,
  2010arXiv:1001.1170
  \bibitem[Ruszkowski etal 2010]{ruszk10} Ruszkowski, M., Lee, D., Bruggen, M.,
  Parrish, I. \& Oh, S. P. 2010,
  2010arXiv:1010.2277
  \bibitem[Ryu etal 2008]{ryu08} Ryu, D., Kang, H., Cho, J. \& Das, S. 2008,
  Science, 320, 909
  \bibitem[Ryu etal 2003]{ryu03} Ryu, D., Kang, H., Hallman, E. \& Jones, T. W. 2003,
  ApJ, 593, 599
  \bibitem[Schekochihin \& Cowley 2006]{sche06} Schekochihin, A. A. \& Cowley, S. C
  2006, Physics of Plasmas, 13, 056501
  \bibitem[Schekochihin etal 2004]{schek04} Schekochihin, A. A., Cowley, S. C., Taylor, S. F.,
  Maron, J. L. \& McWilliams, J. C. 2004, ApJ, 612, 276
  \bibitem[Shuecker etal 2004]{schu04} Schuecker, P., Finoguenov, A., Miniati, F., Bohringer, H.
  \& Briel, U. G. 2004, A\&A, 426, 387
  \bibitem[Spitzer 1978]{sp78} Spitzer, L. 1978, Physical Processes in the Interstellar Medium
  (New York: Wiley)
  \bibitem[Vazza etal 2009]{vazza09} Vazza, F., Brunetti, G., Kritsuk, A., Wagner, R.,
  Gheller, C. \& Norman, M. L. 2009, A\&A, 504, 33
  \bibitem[Waelkens etal 2009]{waelk09} Waelkens, A. H., Schekochihin, A. A. \& Ensslin, T. A.
  2009, MNRAS, 398, 1970
  \bibitem[Xu etal 2010]{xu10} Xu, H., Li, H., Collins, D. C., Li, S. \& Norman, M. L. 2010,
  ApJ, 725, 2152

\end{thebibliography}

\end{document}